\documentstyle[prd,aps]{revtex}
\begin{document}
\draft
 
%
%
\twocolumn[\hsize\textwidth\columnwidth\hsize\csname
@twocolumnfalse\endcsname
 
\preprint{SUSSEX-AST 95/10-1, IEM-FT-95/116, astro-ph/9510029}
\title{Metric Perturbations from Quantum Tunneling in Open Inflation}
\author{Juan Garc\'\i a-Bellido}
\address{Astronomy Centre, University of Sussex, Falmer, Brighton BN1
9QH,\ U.K.}
\date{April 8, 1996}
\maketitle
\begin{abstract}
We study the effect that quantum fluctuations produced during the
nucleation of a single-bubble open inflationary universe have on the
amplitude of temperature anisotropies in the microwave background. We
compute the instanton action for the quantum tunneling between the
false and true vacua in open inflation models and show that the
amplitude of quantum fluctuations of the bubble wall is very sensitive
to the gravitational effects of the true vacuum. We study the spectrum of
quantum fluctuations of the bubble wall and confirm that there is only an
inhomogeneous ($k^2 = -3$) discrete mode associated with transverse
traceless fluctuations of the bubble wall. This super-curvature
mode could in principle
distort the anisotropy of the microwave background. We compute the
amplitude of the gauge invariant metric perturbations induced by the
bubble wall fluctuations on a comoving hypersurface, and calculate the
induced amplitude of temperature fluctuations in the microwave
background, for arbitrary values of $\Omega_0$. We find that in the limit
$\Omega_0\simeq 1$, the quadrupole dominates the angular power
spectrum, like in the usual Grishchuk-Zel'dovich effect. The resulting
bounds on the amplitude of
quantum fluctuations of the bubble wall from the absence of such an
effect in the observed microwave background anisotropies
are quite strong. We also study the
contribution from a discrete long wavelength super-curvature mode
($k^2 \simeq 2m^2/3H^2$) that appears in the spectrum of open de Sitter
vacuum fluctuations. We constrain the parameters of the models of open
inflation so that  these modes do not distort the observed temperature
anisotropy.
\end{abstract}
\pacs{PACS numbers: 98.80.Cq \hspace*{3cm} Preprint SUSSEX-AST 95/10-1,
astro-ph/9510029}
 
\vskip2pc]
 
\section{Introduction}
 
Until recently, one of the most robust predictions of inflation was
the extreme flatness of our local patch of the universe. However, in
the last few months there has been a lot of excitement about the
possibility of producing an open universe from
inflation~\cite{BGT,STY,LM}. An open universe could resolve the age
crisis caused by the observations of a relatively large Hubble
constant, $H_0 = 69 \pm 8$ km/s/Mpc, which corresponds (for a flat
universe without a cosmological constant) to a very small age of the
universe, $t_0 = 9.5 \pm 1.1$ Gyr~\cite{Nature}, in conflict with the
ages of globular clusters, $15.8 \pm 2.1$ Gyr~\cite{GlobClus}. An
alternative solution could be the introduction of a non-zero
cosmological constant which could accommodate both a flat and old
universe with a large expansion rate, but there still remains the
question of why the cosmological constant is so small. But perhaps
the true excitement comes from the fact that open inflation provides a
new way of solving the classical problems of the hot big bang
cosmology, the homogeneity and flatness problems. In standard
inflation the two are intimately related and it is not possible to
relax one (flatness) without affecting the other
(homogeneity)~\cite{KTF}. Open inflation solves the homogeneity
problem by inflating the universe in a false vacuum and then
nucleating a very symmetric bubble within which our universe expanded
to `almost' flatness.
 
The first models of open inflation~\cite{BGT} considered a single
field trapped in a metastable state that later tunneled to the true
vacuum with a non-zero energy density.  The field then rolled down a
very flat potential, inflating the required amount of $e$-folds to
produce an open universe. At the end of inflation the universe
reheated to give the well known hot big bang cosmology. These models
had the unpleasant feature of strongly contrived potentials, since in
order to tunnel without producing too large inhomogeneities a large
mass in the false vacuum is needed, while a very small mass for the
inflaton field is required to give the observed amplitude of density
perturbations in the cosmic microwave background (CMB). Linde and
Mezhlumian~\cite{LM} suggested a simple way out by including two
fields, one with a large mass, responsible for tunneling, and the
other with a very small mass, responsible for inflation in the true
vacuum.
 
According to this picture we live inside a bubble that nucleated from
de Sitter space by quantum tunneling with an extremely small
probability. This ensures, first of all, that there will be no
other nucleating events, at least in our past light cone, and
therefore the initial state is pure de Sitter vacuum. Second,
that the nucleated bubble is extremely spherically symmetric.
Although the homogeneity problem is thus solved at the classical
level, there might still be large quantum fluctuations during the
process of quantum tunneling.
 
There are in principle two sources of metric perturbations in open
inflation, the vacuum fluctuations of the inflaton field that are
stretched to cosmological scales by the expansion, and quantum
fluctuations of the bubble wall produced during bubble nucleation.
The first have been extensively studied in recent
papers~\cite{BGT,STY,YST,HST,LW}; the second have been addressed
by Linde and Mezhlumian~\cite{LM}, and more recently by Hamazaki et
al.~\cite{HST}, for an empty bubble. We study the fluctuations of
the wall when the bubble is not empty. The calculations will be done
in the thin wall approximation, which is valid for most potentials
with a deep false vacuum minimum and a large potential barrier between
the two vacua. Most results follow Coleman-De Luccia's
formalism~\cite{CDL}, valid when the tunneling occurs from de Sitter
to Minkowski space-time. However, the new ingredient in open inflation
is precisely the non-zero energy density of the true vacuum which
could still drive inflation to almost flatness. The instanton action
associated with the more general quantum tunneling process from de
Sitter to de Sitter was computed long ago by Parke~\cite{Parke}. We
will use his results to calculate the tunneling action of open inflation.
We compute the average amplitude of quantum fluctuations
of the bubble wall from variations of the instanton action. Following the
covariant formalism of Garriga and Vilenkin~\cite{GV}, we study
the spectrum of inhomogeneous scalar modes associated with quantum
fluctuations of the bubble wall, and find that there is only a discrete mode,
with $k^2 = -3$.\footnote{$k^2 = 1$ corresponds to the curvature scale.}
This mode could in principle contribute very strongly to the anisotropy
of the CMB. We study its contribution to the CMB in a gauge invariant
way and present the results in the Appendix, both analytically for
$\Omega_0 \simeq 1$ and numerically for $\Omega_0 < 1$.
We analyze the constrains on the open inflation
models from the absence of such and effect in the anisotropies of the
microwave background, as observed by COBE~\cite{COBE}.
It turns out that there are important constraints on the models, but not
enough to rule them out.
 
Another important issue is whether large quantum fluctuations of the
inflaton field before tunneling could propagate inside the bubble and
distort the CMB. This is a relevant question in the case that the
universe in the false vacuum is actually in a process of
self-reproduction, and thus extremely inhomogeneous~\cite{Book}. In
that case, very large scale metric perturbations could affect the
amplitude of the lowest multipoles of the temperature anisotropies in
the background radiation. This is the so-called Grishchuk-Zel'dovich
effect~\cite{GZ}. We have recently evaluated this effect in the open
universe case~\cite{OGZ} and found strong constraints on the amplitude
of very long wavelength perturbations contributing to the lowest CMB
multipoles. In the case of open inflation models where the mass of the
scalar field in the false vacuum is smaller than the Hubble parameter,
there is a discrete super-curvature vacuum mode~\cite{STY}, with $k^2
\simeq 2m^2/3H^2 < 1$, that could in principle distort the CMB, as
discussed by Yamamoto et al. in Ref.\cite{YST}. We derive bounds on
the parameters of open inflation models from the absence of such an
effect in the microwave background anisotropies.
 
\section{Quantum tunneling}
 
In this Section we review the calculation of Parke~\cite{Parke}
on the instanton action for the quantum tunneling between a false and
a true vacuum in de Sitter space. We assume the potential has a
large barrier between the two minima, so that the thin wall
approximation remains valid, and a large mass in the false vacuum. One
of the dangers of quantum tunneling, for a small mass of the tunneling
field, is the existence of the Hawking-Moss instanton~\cite{HM}. In
this case, the field jumps to the top of the barrier between the two
vacua and very slowly `rolls down' the potential. The problem then is
that there are large quantum fluctuations which are not inflated away,
and these large perturbations would unacceptably distort the observed
anisotropy of the CMB. For that reason alone it is assumed that the
mass of the tunneling field should be much larger than the rate of
expansion at the false vacuum.
 
The Euclidean tunneling action for a single scalar field can be
written as
\begin{equation}
S_E = \int d^4x\,\sqrt{-g} \Big[-{1\over2\kappa^2}\,R +
{1\over2} (\partial\phi)^2 + U(\phi)\Big]\,,
\end{equation}
where $\kappa^2 \equiv 8\pi G$ and the Euclidean $O(4)$-invariant
metric is $ds_E^2 = d\tau^2 + a^2(\tau)\,d\Omega_3^2$. The curvature
scalar is given by $R = - 6a^{-2}(aa''+a'^2-1)$, where a prime denotes
derivative with respect to Euclidean time. Integrating by parts
and using the Euclidean equations of motion,
$a'^2 - 1 = a^2\kappa^2\,[\phi'^2/2 - U(\phi)]/3$, we find
\begin{eqnarray}
S_E & = & 2\pi^2 \int d\tau\,\Big[a^3\,\Big({1\over2}\phi'^2 +
U(\phi)\Big) \nonumber\\[1mm]
&&\hspace{1.5cm}
+ {3\over\kappa^2}\,(a^2 a'' + a a'^2 - a) \Big] \nonumber\\[1mm]
& = & - {12\pi^2\over\kappa^2} \int d\tau\,a\,(1-a^2H^2) \,,
\end{eqnarray}
where $H^2 \equiv \kappa^2\,U/3$. The instanton (or bounce) action
which determines the probability of tunneling is given by $B =
S_E(\phi) - S_E(\phi_F)$. We define $U_F \equiv U(\phi_F)$ and
$U_T \equiv U(\phi_T)$ as the false and true vacuum energies,
respectively, which characterize the end points of the quantum
tunneling. Taking into account the contributions to the instanton
action coming from both the wall and the interior of the bubble,
Parke found the following expression for the bounce
action~\cite{Parke},
\begin{eqnarray}\label{bounce}
B(a) & = & 2\pi^2 a^3 S_1 \nonumber\\[2mm]
& + & {4\pi^2\over\kappa^2} \Big[ {1\over H_T^2}
\Big((1 - a^2 H_T^2)^{3/2} - 1\Big) \nonumber\\[2mm]
& - & {1\over H_F^2} \Big((1 - a^2 H_F^2)^{3/2} - 1\Big)\Big] \,,
\end{eqnarray}
where
\begin{equation}\label{wall}
S_1 = \int_{\phi_F}^{\phi_T} d\phi\,[2(U(\phi)- U_F)]^{1/2}
\end{equation}
and $U_F - U_T \equiv \epsilon$. In general we choose $\epsilon
\ll U_T$ but this is not essential. For the thin wall approximation to
be valid we require that the width of the bubble wall, $\Delta a$, be
much smaller than its radius of curvature,
\begin{equation}\label{thin}
{\Delta a\over a} \simeq
{H_T\,(\phi_T - \phi_F)\over[2(U_0 - U_F)]^{1/2}} \ll 1 \,,
\end{equation}
where $U_0$ is the value of the potential at the maximum. The only
requirement is that the barrier between $\phi_F$ and $\phi_T$ be
sufficiently high, i.e. $U_0 \gg U_T$.
 
It is now possible to compute the radius of curvature of the bubble
wall for which the action (\ref{bounce}) is an extremum,
\begin{eqnarray}
B'(a) & = & {12\pi^2 a\over\kappa^2} \Big[ {\kappa^2\over2}\,
S_1\,a \nonumber\\[2mm]
& - & (1 - a^2 H_T^2)^{1/2} + (1 - a^2 H_F^2)^{1/2} \Big] = 0\,.
\end{eqnarray}
An exact solution~\cite{Parke} can be written in terms of
dimensionless parameters $\alpha$ and $\beta$,
\begin{eqnarray}\label{alpha}
a^2 H_T^2 & = & {\alpha^2\over\alpha^2+(1+\alpha^2\beta)^2}  \,,\\[2mm]
\alpha = a_0\,H_T & \equiv & {3S_1\over\epsilon}\,H_T \,, \hspace{1cm}
\beta = {\epsilon\over4U_T} \,.
\end{eqnarray}
The parameter $\alpha$ characterizes the strength of the
gravi\-tational interaction in the true vacuum. The extremal
solution (\ref{alpha}) is valid both in the limit $\alpha^2 \ll 1$,
for which we recover the usual tunneling result, $a = a_0$, from de
Sitter to Minkowski ($H_T=0$); and in the limit $\alpha^2\beta
\gg 1$, which gives $a = 4/(\kappa^2\,S_1)$. In both cases the radius
of curvature satisfies $a \ll H_T^{-1}$. On the other hand, the
largest radius of curvature occurs for $\alpha^2\beta = 1$, that is
$a = H_F^{-1}$.
 
The extremal action corresponds to the $O(3,1)$ symmetric bubble. We
are interested in deviations from perfect isotropy and homogeneity,
i.e. on the quantum fluctuations generated during bubble nucleation.
Linde and Mezhlumian~\cite{LM} evaluated a typical quantum deviation
of the radius of curvature of the bubble by computing the first
quantum correction to the tunneling action, $S = S_0 + \hbar \,
\Delta S$, where $\Delta S = B''(a)\,(\delta a)^2/2$, and
the second derivative of the bounce action (\ref{bounce}) at the
extremum is exactly given by
\begin{eqnarray}
B''(a) & = & {12\pi^2\over\kappa^2} \Big[ (1 - a^2 H_T^2)^{-1/2}
- (1 - a^2 H_F^2)^{-1/2} \Big] \nonumber\\[2mm]
& = & -\,18\pi^2\,{S_1^2\over\epsilon}\ {[\alpha^2 +
(1+\alpha^2\beta)^2]^{1/2}\over
(1+\alpha^2\beta)(1-\alpha^2\beta)} \nonumber\\[2mm]
& \simeq & -\,18\pi^2\,{S_1^2\over\epsilon}\ (1-\alpha^2\beta)^{-1}  \,.
\end{eqnarray}
In order to evaluate a typical deviation of the curvature of the bubble,
we can estimate $\Delta S \sim1$, see Ref.~\cite{LM}, and thus,
\begin{equation}\label{delta}
\delta a \simeq {\sqrt\epsilon\over 3\pi S_1} \,
\left|1-\alpha^2\beta\right|^{1/2} \,.
\end{equation}
In the limit $\alpha^2\beta \ll 1$ we recover the results of
Refs.~\cite{GV,LM}. On the other hand, the last factor could strongly
affect the overall curvature of the nucleated bubble, when
gravitational effects are important. For instance, for $\alpha^2\beta
= 1 \ (a = H_F^{-1})$, the extremum solution (\ref{alpha}) is exact
and there are no quantum fluctuations in the curvature of the bubble.
 
We are actually interested in the spectrum of
inhomogeneous quantum fluctuations of the bubble wall, which would
appear from the inclusion of gradient terms in the bounce action.
These inhomogeneous scalar modes were studied by Garriga and
Vilenkin~\cite{GV} for an empty bubble, using a covariant formalism in
an embedding de Sitter space. In the context of open inflation models,
the bubble is not empty and the radius of curvature of the bubble at
the moment of nucleation is smaller than the de Sitter horizon scale,
$H_F^{-1}$. The geometry of the three-dimensional bubble is
characterized by its extrinsic and intrinsic curvature. The
unperturbed bubble world-sheet has an induced metric $\,g_{ab} =
\partial_a x^\mu\,\partial_b x_\mu\,$, where the subindices $\{a,b\}$
label coordinates on the bubble wall, and $\{\mu,\nu\}$ label
space-time coordinates. The extrinsic curvature of the bubble wall is
$\,K_{ab} = - (\dot a/a)\,g_{ab}$. However, as the bubble expands,
$\dot a/a = H\,\coth(Ht) \to H$, the curvature scale of the bubble
approaches the horizon scale $H^{-1}$ and remains fixed. Here $H$
stands for the Hubble constant of the embedding de Sitter space.
 
We are interested in perturbations in the space-time coordinates of
the bubble wall. Since only motion transverse to the bubble wall is
physically observable (the rest can be eliminated by a coordinate
transformation), we will only consider linear perturbations of the
type $\,\delta x^\mu = \varphi\,n^\mu$, where $\varphi$ is a scalar
that characterizes the fluctuations normal to the surface ($n^\mu =
H\,x^\mu$). The
metric perturbations become $\delta g_{ab} = - 2 \varphi\,K_{ab} +
\partial_a\varphi \partial_b\varphi + \varphi^2 K_a^{\ c}K_{cb} \simeq
- 2 \varphi\,K_{ab}$, to first order. The equation of motion for the
scalar fluctuation $\varphi$ can be obtained from the variation of the
extrinsic curvature scalar ($K = - 3H =$ constant), $\delta K =
\nabla^2\varphi + K_{ab} K^{ab} \varphi - K_{ab}
\partial^a\varphi\partial^b\varphi - \varphi^2 K_{ab} K^{ac}
K_c^{\ b} \simeq \nabla^2\varphi + K_{ab} K^{ab} \varphi = 0$, to
first order,\footnote{Note that the full expression $\delta K =0$ still
corresponds a mode with $k^2=-3$.}
\begin{equation}\label{scalar}
\delta K = \nabla^2\varphi + 3H^2 \varphi =
{k^2 + 3\over a^2}\,\varphi = 0 \,.
\end{equation}
The bubble wall fluctuations thus correspond to an inhomogeneous
scalar mode characterized by $k^2 = -3$, with the peculiar property
that the associated curvature perturbation is transverse
traceless~\cite{GV,HST},
\begin{equation}\label{Rab}
\delta\overline R^{(3)}_{ab} = - H\,\delta\overline K_{ab} = - H
(\nabla_a\nabla_b \varphi + H^2 g_{ab}\,\varphi)\,,
\end{equation}
while the Ricci scalar remains unperturbed,
\begin{equation}\label{delR}
\delta R^{(3)} = - H\,{k^2 + 3\over a^2}\,\varphi = 0\,.
\end{equation}
 
In principle there could have been other inhomogeneous modes at bubble
nucleation, but the fact that the bubble wall asymptotically acquires
a fixed curvature determines that only the inhomogeneous scalar mode
with $k^2 = -3$ survives on the surface of the bubble.
 
\section{Metric perturbations}
 
We now study the effect that quantum fluctuations of the bubble wall
produce on the microwave background. In order to do this, we have to
relate the metric perturbations in the (2+1)-dimensional bubble wall,
at a fixed radial coordinate, with metric perturbations on a
3-dimensional comoving equal-time hypersurface inside the bubble. For
that purpose, we recall the open de Sitter coordinates of
Ref.~\cite{BGT}: Region~I contains the interior of the bubble and is
parametrized (in units of $H^{-1}$) by $\,ds^2 = - d\zeta^2 +
\sinh^2\zeta \,(d\xi^2 + \sinh^2\xi\, d\Omega^2)$ with coordinates
$(\zeta,\xi)$, while Region~II is outside the bubble and is described
by the metric $\,ds^2 = d\sigma^2 +
\sin^2\sigma \,(- d\tau^2 + \cosh^2\tau\,d\Omega^2)$ with coordinates
$(\tau,\sigma)$. In these coordinates the bubble wall is a time-like
hypersurface at a fixed coordinate $\sigma$ in Region~II, which can be
analytically continued into a space-like hypersurface at a fixed
comoving time $\zeta = i\,\sigma$ inside the bubble. Thus perturbations
in the bubble wall hypersurface propagate inside as metric perturbations
in a comoving equal-time hypersurface. The (2+1)-dimensional $k^2 = -3$
mode of Ref.~\cite{GV} corresponds analytically to the 3-dimensional
open universe discrete mode with $k^2 = -3$ discussed by Hamazaki et
al. \cite{HST}.
 
We want to evaluate, in linear perturbation theory, the primordial
metric perturbations associated with these quantum fluctuations. The most
general scalar metric perturbations can be written
as~\cite{B80,KS84,MFB92}
\begin{eqnarray}\label{pertbn}
ds^2 & = & a^2(\eta)\Big[- (1+2A) d\eta^2 + 2B_{|i}\,dx^id\eta
\nonumber \\ [2mm]
& + & \Big\{ (1+2{\cal R}) \gamma_{ij} + 2E_{|ij} \Big\} dx^i dx^j
\Big] \,,
\end{eqnarray}
where $\{i,j\}$ label the 3-dimensional open space coordinates with
metric $\gamma_{ij}$. The four linear scalar perturbations are not
independent. Under a gauge transformation $\,\tilde\eta = \eta +
\xi^0(\eta,x^k)$, $\,\tilde x^i = x^i + \gamma^{ij} \xi_{|j}
(\eta,x^k)$, the metric perturbations transform as
\begin{eqnarray}\label{trans1}
\tilde A & = & A - {\xi^0}' - {a'\over a}\xi^0 \,; \hspace{1cm}
\tilde{\cal R} = {\cal R} - {a'\over a}\xi^0 \,,\\ [2mm]\label{trans2}
\tilde B & = & B + \xi^0 - \xi' \,; \hspace{1.5cm}
\tilde E = E - \xi \,,
\end{eqnarray}
where a prime denotes derivative with respect to conformal time
$\eta$. There are however only two independent gauge invariant
gravitational potentials~\cite{MFB92},
\begin{eqnarray}
\Phi & = & A + {1\over a}\,\Big[a(B-E')\Big]'\,,\label{GI}\\ [2mm]
\Psi & = & {\cal R} + {a'\over a}\,(B-E')\,,
\end{eqnarray}
which are further related through the perturbed Einstein equations,
\begin{eqnarray}\label{EEQ}
\Phi + \Psi & = & 0 \,,\\ [2mm]
4\,{k^2 + 3\over a^2}\,\Psi & = & 2\,\delta\rho\,.
\end{eqnarray}
Here $\delta\rho$ is the gauge invariant density perturbation~\cite{B80}.
Note that for the $k^2=-3$ mode of bubble wall fluctuations, the amplitude
of density perturbations is identically zero. This is a very special mode,
as was first pointed out by Lifshitz and Khalatnikov~\cite{LK}. We will
study in detail its effect on metric perturbations.
 
The scalar metric perturbations can be separated into $A(\eta, x^i) =
A(\eta)Q(x^i)$,\footnote{From now on $A$, $B$, etc. stand for the
$\eta$-dependent functions.} where $Q(x^i)$ are the scalar harmonics
of a spatially open universe, solutions of the Helmholtz
equation~\cite{Harrison},
\begin{equation}\label{harmonics}
L^2 Q(\xi,\Omega) = - k^2\,Q(\xi,\Omega)\,,
\end{equation}
where $L^2 = \sinh^{-2}\xi\ \partial_\xi(\sinh^2\xi\ \partial_\xi) +
\sinh^{-2}\xi\ L^2_\Omega$ is the open universe Laplacian. These
solutions have the general form $Q_{klm}(\xi,\Omega) = \Pi_{kl}(\xi)\,
Y_{lm}(\theta,\phi)$, see the Appendix. The scalar
harmonics can be used to construct a traceless tensor
\begin{equation}\label{antisym}
Q_{ij} = {1\over k^2}\,Q_{|ij} + {1\over3}\,\gamma_{ij}\,Q\,,
\end{equation}
satisfying $Q_i^{\ i} = 0$, see Eq.~(\ref{harmonics}), as well as
\begin{equation}\label{anti}
\nabla^i Q_{ij} = -\,{2\over3}\,{k^2+3\over k^2}\,\nabla_jQ\,.
\end{equation}
 
We can now investigate the contribution of the discrete $k^2 = -3$
mode to the primordial perturbations in an open universe. In principle
it is possible to analyze the amplitude of metric perturbations on any
hypersurface~\cite{B80}. However, we believe it is most convenient
to study them on comoving hypersurfaces, where they have a clear
physical meaning as curvature perturbations. Furthermore, the
perturbed bubble wall hypersurface has the property of being
also a uniform-expansion ($\delta K = 0$) hypersurface, see
Eq.~(\ref{scalar}). It turns out that, for $k^2 = -3$
(and only for this mode), the same gauge transformation that takes
you  to a comoving hypersurface,  also takes you to a
uniform-Hubble-constant hypersurface~\cite{B80},
\begin{eqnarray}\label{trace}
\delta K &=& - {3\over a} \Big[{\cal R}' - {a'\over a}
A + {k^2\over3} (B - E')\Big] Q \equiv 0\,, \\ [2mm]
\delta\overline K_{ij} &=& - {k^2\over a}\,(B - E')\,Q_{ij} \equiv
{3\over a} \Big({\cal R}' - {a'\over a} A\Big) Q_{ij} \,,
\end{eqnarray}
with intrinsic curvature separated into its trace and traceless
parts,
\begin{eqnarray}\label{traceless}
\delta R^{(3)} & = & 4\,{k^2 + 3\over a^2}\,{\cal R}\,Q \equiv 0\,,
 \\ [2mm]
\delta\overline R^{(3)}_{ij} & = & - {k^2\over a^2}\,{\cal R}\,Q_{ij}\,.
\end{eqnarray}
Furthermore, for $k^2 = -3$, the dynamic equations can be written
as~\cite{B80}
\begin{eqnarray}\label{EM}
A & = & - \,{w\over1+w}\, \eta\,,\\[2mm]
\Big[a^2 ({\cal R}' - {a'\over a} A)\Big]'
& = & {k^2\over3} ({\cal R} + A)\,a^2\,,
\end{eqnarray}
where $w=p/\rho$, and $\eta = (\delta p - \delta\rho\,dp/d\rho)\,Q$
is the gauge invariant non-adiabatic part of matter perturbations,
which vanishes for the single-field (adiabatic) perturbations of open
inflation. In that case, the remaining scalar perturbation satisfies the
equation
\begin{equation}\label{eqR}
{\cal R}'' + 2{a'\over a}\,{\cal R}' + {\cal R} = 0\,.
\end{equation}
 
During inflation, $a(\eta) = a_0/\sinh\eta$, there is an exact solution
to this equation given by
\begin{equation}\label{Rinf}
{\cal R} = C\,{a'\over a^2} = C\,{\dot a\over a}  \,,
\end{equation}
where $C = \delta a$ is the amplitude of quantum fluctuations
of the bubble wall, see Eq.~(\ref{delta}). It is easy to see that
during inflation ${\cal R}$ becomes constant,
\begin{equation}\label{Rconst}
{\cal R}_0 = H \delta a \,.
\end{equation}
This curvature perturbation is a very peculiar one. Let us evaluate the
gauge invariant potentials $\Phi$ and $\Psi$ in Eq.~(\ref{GI}). Substituting
the solution (\ref{Rinf}), together with ${\cal R}'= - C/a$, we find
\begin{eqnarray}\label{PhPs}
\Phi & = & {1\over a}\,(a {\cal R}')' = 0\,,\\[2mm]
\Psi & = & {\cal R} +  {a'\over a}\,{\cal R}' = 0\,.
\end{eqnarray}
Therefore, this mode has vanishing gauge invariant potentials, as well as
vanishing gauge invariant density perturbations.
One might be tempted to dismiss this mode
altogether; however, it is possible to see that the traceless parts
of the extrinsic and intrinsic curvatures do not vanish,
\begin{eqnarray}\label{curva}
\delta\overline K_{ij}  &=& {3\over a}\,{\cal R}'\,Q_{ij} =
-\,{3\over a^2}\,C\,Q_{ij} \,,\\[2mm]
\delta\overline R^{(3)}_{ij}  & = & {3\over a^2}\,{\cal R}\,Q_{ij} =
{3\over a^2}\,C\,{a'\over a^2}\,Q_{ij}\,.
\end{eqnarray}
Therefore, this $k^2=-3$ mode can be understood as a transverse
traceless perturbation, see Eqs.~(\ref{antisym}, \ref{anti}). This was
pointed out in Ref.~\cite{HST,Jaume}.
During inflation, $a'/a^2 = \dot a/a \to H =$ constant, and thus
$\delta\overline R^{(3)}_{ij} = - H\,\delta\overline K_{ij} $, which
coincides with the bubble wall fluctuations' relation (\ref{Rab}),
as expected.
 
Spatial curvature, $a^{-2} = H^2(1-\Omega)$, will vanish during the
last stages of inflation and will be negligible ($\Omega \simeq 1$) during
radiation. It is then easy to show with Eq.~(\ref{eqR})
that ${\cal R}$ remains approximately constant. However, as we enter the
matter era, $a(\eta) = a_0 (\cosh\eta - 1)$, spatial curvature will become
important, in order to produce an open universe ($\Omega < 1$).
As a consequence, ${\cal R}$ is no longer a constant and evolves
with Eq.~(\ref{eqR}). There is an exact solution to this equation
during the matter era,
\begin{equation}\label{Rmat}
{\cal R} = {\cal R}_0\,G(\eta) \equiv  {\cal R}_0\,3\,{\eta\sinh\eta -
2(\cosh\eta-1)\over(\cosh\eta -1)^2}\,,
\end{equation}
satisfying $G(0)=1$. We are interested however in the gauge invariant
gravitational potential (\ref{GI}),
\begin{equation}\label{Phim}
\Phi = - ({\cal R} + {a'\over a}\,{\cal R}') =  -\,{3\over5}\,{\cal R}_0\,
F(\eta)\,,
\end{equation}
where
\begin{equation}\label{Feta}
F(\eta) \equiv  5\,{\sinh^2\eta-3\eta\sinh\eta +4(\cosh\eta-1)\over
(\cosh\eta -1)^3}\,,
\end{equation}
satisfying $F(0)=1$.
 
In the next section we will evaluate the amplitude of temperature
anisotropies from all possible modes present in open inflation.
These include the $k^2=-3$ mode associated with bubble wall
fluctuations, as well as the discrete super-curvature mode with
$k^2=2m^2/3H^2$ present in the spectrum of open de Sitter vacuum
fluctuations. There is also the continuum of sub-curvature modes
associated with quantum fluctuations of the inflaton field during
the second stage of inflation.
 
\section{Temperature anisotropies}
 
In this Section we study the constraints that temperature anisotropies
in the CMB impose on the models of open inflation. For any model, the
value of $\Omega_0$ today depends very critically upon the number of
$e$-folds of inflation from the tunneling event to the end of
inflation. For $N_e = 65$ we find $\Omega_0$ very close to one. In
fact, since
\begin{equation}\label{Omega0}
|1-\Omega_0| \sim \exp(-2N_e) \times 10^{56}\,,
\end{equation}
a few $e$-folds of inflation less will produce a wide open universe. In
most models, the second stage of inflation within the bubble occurs
in the usual way with a very flat scalar potential, where 65 $e$-folds
correspond to a value of the inflaton field $\sigma \simeq 3\,M_{\rm
P}$. In that case, tunneling to a value of the inflaton just below
$3\,M_{\rm P}$ would produce an open universe~\cite{LM}. In the
single-field models of open inflation proposed in Ref.~\cite{BGT,STY},
the tunneling field is also the inflaton field. As we discussed in
Section 2, in order that the field does not tunnel to the top of the
potential, and thus produce large amplitude metric perturbations, we
need a mass at the false vacuum which is larger than the rate of
expansion, $M_F \gg H_F$. This condition is enough to suppress the
amplitude of metric perturbations before tunneling. However, in the
two-field models of Ref.~\cite{LM} the tunneling field $\phi$ could be
very heavy while the inflaton field $\sigma$ has a small mass, both in
the false and the true vacuum. In this case, Sasaki et al. have shown
that there exists a discrete super-curvature mode in the spectrum of
vacuum fluctuations in the open de Sitter space~\cite{STY}. This mode
could in principle affect the lowest multipoles of the temperature
anisotropy of the CMB, in what is known as the Grishchuk-Zel'dovich
effect, studied in Ref.~\cite{OGZ} for an open universe.
 
The dominant effect on large scales is known as the Sachs-Wolfe effect.
Due to this effect, metric perturbations on the surface of last
scattering are responsible for temperature fluctuations in the
CMB~\cite{Liddle},
\begin{equation}\label{SWE}
{\delta T\over T} = {1\over3}\,\Phi(0) + 2 \int_0^{\eta_0}
\Phi'(\eta_0-r)\,dr\,,
\end{equation}
where $\eta_0 = \cosh^{-1}(2/\Omega_0 - 1)$ is the distance to the
last scattering surface. For $\Omega_0 < 2/(1+\cosh1) \simeq 0.786$,
the surface of last scattering is located  beyond the curvature scale.
 
We can expand the observed temperature anisotropies in terms of
spherical harmonics,
\begin{equation}\label{DTT}
{\delta T\over T}(\theta,\phi) = \sum_{l,m}\,a_{lm}\,
Y_{lm}(\theta,\phi)\,,
\end{equation}
and evaluate the angular power spectrum
$\,C_l \equiv \langle|a_{lm}|^2\rangle$, defined as the
ensemble average of the $l$-th multipole of the CMB temperature
anisotropy. Observations in fact suggest that
\begin{equation}\label{obs}
l(l+1)\,C_l \lesssim {24\pi\over5}\,{Q_{\rm rms}^2 \over T_0^2}
\simeq 8\times 10^{-10}\,,
\end{equation}
for the lowest multipoles of the CMB anisotropies, where $Q_{\rm rms}
\simeq 20\,\mu$K~\cite{Gorski}.
 
The contribution of the different metric perturbation modes to the
multipole components of the angular power spectrum can be separated
into a discrete part, which includes the bubble wall fluctuations,
labeled by $\nu=2$; the super-curvature mode $\,k^2_{\rm VL}
\simeq 2m_F^2/3H_F^2 < 1\,$ of open de Sitter vacuum~\cite{STY},
plus a continuum of sub-curvature modes from
fluctuations of the inflaton field in the last stage of inflation,
\begin{eqnarray}\label{LCL}
l(l+1)C_l & = & 2\pi^2\,l(l+1) \,\langle{\cal R}_0^2\rangle_{\rm wall}
I_{\nu l}^2 \nonumber\\[2mm]
& + & 2\pi^2\,l(l+1) \,N_l^2\,B_l^2\,k^2_{\rm VL}\,
\langle{\cal R}^2\rangle_{\rm VL}\nonumber\\[2mm]
& + & 2\pi^2\,l(l+1)
\int_1^\infty {dk\over k}\,{\cal P}_{\cal R}(k) I_{kl}^2\,.
\end{eqnarray}
Here ${\cal P}_{\cal R}(k)$ is the spectrum of primordial curvature
perturbations. For the $\nu=2$ mode associated with the bubble wall,
$\langle{\cal R}_0^2\rangle_{\rm wall}$ is the average square
amplitude of the curvature perturbation (\ref{Rconst}). The ``window
function'' $I_{kl}^2$ indicates how a given scale contributes to the
$C_l$'s. We have evaluated $I_{\nu l}^2$ for the $\nu=2$ mode in the
Appendix. The window function $N_l^2\,B_l^2$
associated with the very long wavelength
super-curvature mode $k^2_{\rm VL}$ was computed in Ref.~\cite{OGZ}.
 
We analyze now the different sources of metric
perturbations in models of open inflation and the contraints
that observations of the microwave background impose on the
models.
 
\subsection*{The sub-curvature modes}
 
Let us consider the temperature anisotropies produced by the continuum
spectrum of sub-curvature modes, see also Ref.~\cite{YST}. The
amplitude of scalar metric perturbations produced by quantum
fluctuations of the inflaton field $\sigma$ during the second phase of
inflation is approximately given by~\cite{LW}
\begin{equation}\label{Rinflaton}
{\cal R} = - H\,{\delta\sigma\over\dot\sigma} \,,
\end{equation}
which gives a nearly scale invariant spectrum,
\begin{equation}\label{dens}
\langle{\cal R}^2\rangle^{1/2} =
{4V\over\sigma V'}\,{H \sigma\over M_{\rm P}^2}
\simeq {6 H\over M_{\rm P}}\,.
\end{equation}
 
The observed temperature anisotropies of the microwave background at
large scales~\cite{COBE} are consistent with such a spectrum of metric
perturbations,
\begin{equation}
l(l+1)\,C_l = {2\pi\over25}\,\langle{\cal R}^2\rangle \simeq
{\rm constant} \,.
\end{equation}
Assuming that the observed anisotropies (\ref{obs}) are due solely to
quantum fluctuations of the inflaton requires $H/M_{\rm P} \simeq
10^{-5}$. This is a very general constraint. In the case of a
massive inflaton, this bound requires its mass in the true vacuum
to be
\begin{equation}\label{MT}
{m_T\over M_{\rm P}} \simeq 2\times 10^{-6}\,.
\end{equation}
All models of open inflation should satisfy this constraint.
Let us now discuss the more interesting discrete modes.
 
\subsection*{The $k^2 = -3$ mode.}
 
This mode is associated with quantum fluctuations
of the bubble wall at tunneling that survive as scalar perturbations
after the bubble has acquired a fixed comoving curvature. In order to
evaluate its amplitude, let us parametrize the tunneling potential of
Section II by
\begin{equation}\label{poten}
U(\phi) = U_F + {\lambda\over4}\,\phi^2(\phi-\phi_0)^2 -
\epsilon \Big({\phi\over\phi_0}\Big)^4\,.
\end{equation}
The two minima occur at $\phi_F = 0$ and $\phi_T \simeq \phi_0 \equiv
M\sqrt{2/\lambda}$, and the maximum of the potential is at $U_0 =
\lambda\,\phi_0^4/64\,= M^4/16\lambda$. Following Ref.~\cite{LM}, we
will define $\epsilon \equiv \mu\,U_0$, with $\mu \ll 1$ for the thin
wall approximation to be valid.
 
It is now possible to evaluate the instanton contribution from the
bubble wall (\ref{wall}),
\begin{equation}\label{S1}
S_1 = {M^3\over3\lambda}\,\Big(1 + {11\over32}\,{\epsilon\over
U_0}\Big) \simeq {M^3\over3\lambda}\,,
\end{equation}
while $\alpha$ and $\beta$ are, see Eq.~(\ref{alpha}),
\begin{eqnarray}\label{alfa}
\alpha = {16\over\mu}\,{H\over M} \,, &&\hspace{1cm}
\beta = {\pi\over24}\,{\mu\over\lambda}\,{M^4\over
M_{\rm P}^2\,H^2} \,,\\[2mm]
\alpha^2\beta & = & {32\pi\over3\mu\lambda}\,
{M^2\over M_{\rm P}^2} \,.
\end{eqnarray}
 
The average amplitude of the metric perturbation produced by
quantum fluctuations in the radius of curvature of the bubble at
nucleation, ${\cal R}_0 = H \delta a$, see Eq.~(\ref{Rconst}), can
then be written as
\begin{eqnarray}\label{dwall}
\langle{\cal R}_0^2\rangle^{1/2}_{\rm wall} & \simeq &
{\sqrt{\mu\lambda}\over4\pi}\,{H\over M}\,
\left|1-\alpha^2\beta\right|^{1/2} \nonumber\\[2mm]
& = & \sqrt{2\over3\pi}\,{H\over M_{\rm P}}\,\left|{1-\alpha^2\beta
\over\alpha^2\beta}\right|^{1/2} \,.
\end{eqnarray}
Note that we recover the result of Ref.~\cite{LM} in the limit
$\alpha^2\beta \ll 1$. The constraints on $\langle{\cal R}^2_0
\rangle_{\rm wall}$ from the CMB temperature fluctuations
are discussed in the Appendix, see Eqs.~(\ref{bound1},
\ref{bound2}). Using (\ref{dwall}) we find,
\begin{eqnarray}\label{Rbound}
\left|{1-\alpha^2\beta\over\alpha^2\beta}\right| &\lesssim&
18\pi^2 \,,\hspace{1.8cm} 0.1 \lesssim \Omega_0
\lesssim 0.4\,,\\[2mm]
\left|{1-\alpha^2\beta\over\alpha^2\beta}\right| &\lesssim&
{54\pi^2\over(1-\Omega_0)^2}\,,\hspace{1.1cm}
0.4 \lesssim \Omega_0 \leq 1\,.
\end{eqnarray}
We can now bound the parameters of the model (\ref{alpha}),
\begin{eqnarray}
\mu\lambda &\lesssim&192\pi^3 {M^2\over M_{\rm P}^2}\,,
\hspace{1.6cm} 0.1 \lesssim \Omega_0 \lesssim 0.4\,,\\[2mm]
\mu\lambda &\lesssim& {576\pi^3\over(1-\Omega_0)^2}\,
{M^2\over M_{\rm P}^2}\,,\hspace{1.1cm}
0.4 \lesssim \Omega_0 \leq 1\,.
\end{eqnarray}
Let us give some values to the parameters. Suppose that
$M\sim 10^{-3} M_{\rm P}$, see Ref.~\cite{LM}, and take e.g.
$\Omega_0 \simeq 0.3$. Then
\begin{equation}
\mu\lambda < 6 \times10^{-3} \,.
\end{equation}
This is a relatively weak bound on $\lambda$ for small values
of $\mu$. On the other hand, for $\Omega_0=0.8$, the bound
becomes $\mu\lambda < 0.4$, much weaker and thus much
easier to satisfy. Furthermore, as $\Omega_0$ tends to one,
the bound disappears altogether.
 
In summary, the bound on the parameters of open inflation
models from quantum fluctuations of the bubble wall depends
significantly on the value of $\Omega_0$. In most cases, the
bound is not very strong and it is possible to find models satisfying
the constraints.
 
\subsection*{The discrete vacuum mode.}
 
For open inflation models in which one of the scalar
fields has a mass in the false vacuum much smaller than the Hubble
rate of expansion, there exists a discrete super-curvature mode in
the spectrum of open de Sitter vacuum fluctuations~\cite{STY},
\begin{equation}\label{super} k^2_{\rm VL} = 1 - \Big[\Big({9\over4} -
{m_F^2\over H_F^2}\Big)^{1/2} - {1\over2}\Big]^2 \simeq \ {2\over3}
{m_F^2\over H_F^2}\,, \end{equation}
that propagates inside the bubble. This is the case of one of the
two-field models of Ref.~\cite{LM}, where the inflaton in the false
vacuum is light. The metric perturbation for this mode is
\begin{equation}\label{curv}
\langle{\cal R}^2\rangle_{\rm VL} \simeq \left(
{4 V_F(\sigma)\over V'_F(\sigma)}\,{H_F\over M_{\rm P}^2}\right)^2
\simeq 36\,{H_F^2\over M_{\rm P}^2}\,,
\end{equation}
which could in principle affect the anisotropy of the microwave
background, due to a large value of $H_F$.
 
The Grishchuk-Zel'dovich effect gives the contribution to the
microwave background anisotropy from a very large scale metric
perturbation. In an open universe, one can constrain the amplitude
of a discrete very long wavelength super-curvature mode
$0 < k^2_{\rm VL} < 1$ to satisfy~\cite{OGZ}
\begin{equation}\label{openGZ}
k^2_{\rm VL}\,\langle{\cal R}^2\rangle_{\rm VL} \lesssim
4 \times 10^{-8}\,,
\end{equation}
in the range $\,0.25 \lesssim \Omega_0 \lesssim 0.8$, in order not to
affect the observed low multi\-pole aniso\-tropies of the microwave
background, see Eq.~(\ref{obs}). The constraint is somewhat weaker,
$\,k^2_{\rm VL}\,\langle{\cal R}^2\rangle_{\rm VL} <
2 \times 10^{-9}\,
(1-\Omega_0)^{-2}$, for $\,0.8 \lesssim \Omega_0 < 1$, see
Ref.~\cite{OGZ}.
 
Using the bound (\ref{openGZ}), we find
a relatively weak bound on the allowed mass of the inflaton field in
the false vacuum,
\begin{equation}\label{MF}
{m_F\over M_{\rm P}} \lesssim 4 \times 10^{-5}\,,
\end{equation}
which allows a mass somewhat larger than that in the true
vacuum (\ref{MT}).
 
Another possibility is to chose the inflaton field to be massless in
the false vacuum and thus the discrete mode (\ref{super}) would
correspond to a homogeneous ($k^2 = 0$) mode, which does not
affect the CMB, see Ref.~\cite{OGZ}.
This is the case of hybrid open inflation~\cite{LM}. One could
also have a very massive inflaton field in the false vacuum, as in the
``supernatural'' open inflation model of Ref.~\cite{LM}, in which case
there is no discrete super-curvature vacuum mode.
 
In summary, either the scalar fields are very massive in the false
vacuum, such that no super-curvature modes are present in the spectrum
of metric perturbations, or they are very light so that their
associated super-curvature mode does not distort the observed
anisotropy of the microwave background.
 
\section{Conclusions}
 
In this paper we have computed the amplitude of metric perturbations
produced by quantum fluctuations of the bubble wall at the moment of
tunneling. These could in principle be a source of temperature
fluctuations in the microwave background, in the context of the
present models of single-bubble open inflation. By taking into account
the corrections due to gravitational effects at the tunneling event,
we have found that a non-zero energy density in the true vacuum could
strongly modify the amplitude of fluctuations in the bubble wall.
These fluctuations can be understood as discrete long wavelength modes
associated with perturbations in the curvature of the bubble.
In the open
de Sitter coordinates, the bubble wall is a time-like hypersurface at
a fixed radial coordinate which asymptotically determines a space-like
hypersurface at a fixed comoving time inside the bubble. Small
fluctuations in the curvature of the bubble propagate inside as
perturbations in the time it takes to end inflation~\cite{GP}, and
thus generate metric perturbations on comoving hypersurfaces. However,
quantum fluctuations of the bubble wall generate only
a discrete inhomogeneous ($k^2 = -3$) mode~\cite{GV,HST}.
It is possible to calculate the effect that this transverse traceless
scalar metric perturbation produces on the microwave background.
We have computed this effect for arbitrary values of $\Omega_0$,
and described the results in the Appendix. See also Ref.~\cite{Jaume}
for the limit $\Omega_0\simeq 1$. In section IV we constrain the
parameters of open inflation models to avoid distorsions in the
observed temperature anisotropies. The resulting bounds on the
amplitude of bubble wall quantum fluctuations are quite severe,
although not enough to rule out these models.
 
Furthermore, in the single-field models of Refs.~\cite{BGT,STY}, the
inflaton potential is fine-tuned so that the field has a very large
mass in the false vacuum. This ensures not only that the tunneling
occurs in the thin wall approximation and not along the Hawking-Moss
instanton, but also that there are no super-curvature modes that
propagate inside the bubble~\cite{HST}. However, in the two-field
models of Ref.~\cite{LM} the tunneling field has a very large mass in
the false vacuum, but the inflaton field does not (except in the
``supernatural inflation'' model). This implies two things, first that
there is a discrete super-curvature vacuum mode~\cite{STY} that
propagates
inside the bubble, and second that the amplitude of such a long
wavelength perturbation could be rather large. These two features could
in principle be enough to destroy the isotropy of the CMB to a level
incompatible with observations, see Ref.~\cite{YST}. Using the bounds
on the amplitude and wavelength of such a perturbation from the open
universe Grishchuk-Zel'dovich effect~\cite{OGZ}, we find an upper
bound on the mass of the inflaton field at the false vacuum which is
easily satisfied by all models.
 
As a consequence, the present models
of open inflation seem to work well with very reasonable parameters,
at least as reasonable as those of standard inflation. A different
issue is whether these models will turn out to be the correct
description of the origin of our patch of the universe.
 As mentioned in the introduction, another possible solution to the age
crisis could be that the universe is flat with a non-vanishing
cosmological constant. Fortunately cosmology has become a science and
within a few years we will be able to tell, from the shape and
amplitude of the spectrum of temperature fluctuations, whether our patch
of the universe is indeed open or flat~\cite{Kamion}. A different
possibility is that the present observations of the Hubble parameter
turn out to be wrong and the actual value is well within the range
allowed by a flat universe without a cosmological constant.
 
In any case, it is encouraging to see that the inflationary paradigm
is able to accommodate an open universe, even if we never have to make
use of it. Much more difficult would be to compute a probability
distribution for the value of $\,\Omega_0$.  Such an attempt was made
in Refs.~\cite{LM} and~\cite{Luca}. We believe the problem of
probability measure in cosmology is not yet settled and we still have
to learn how to pose the appropriate questions.

\section*{Acknowledgements}
The author thanks Jaume Garriga, Andrew Liddle, Andrei Linde,
David Lyth and David Wands for many fruitful discussions, and
very specially Misao Sasaki for very generous comments on
the correct normalization of the supercurvature modes.
He also thanks the Aspen Center for Physics for a wonderful atmosphere and
the Physics Department at Stanford, where part of this work developed.
The author is supported by PPARC (U.K.) and a NATO Collaborative Grant.

\appendix
\section*{CMB temperature fluctuations from a $k^2=-3$ mode}
 
The open universe scalar harmonics can be written as
$Q_{klm}(\xi,\Omega) = \Pi_{kl}(\xi)\,Y_{lm}(\theta,\phi)$,
with $Y_{lm}(\theta,\phi)$ the usual spherical harmonics and
\begin{eqnarray}\label{pikl}
\Pi_{kl}(\xi) &=& N_{kl} \
{P^{-l-1/2}_{\nu-1/2}(\cosh\xi)\over\sqrt{\sinh\xi}}\,,
\\[2mm]
N_{kl} &=& \left(\Gamma(l+1+\nu)\Gamma(l+1-\nu)\over2\right)^{1/2}\,,
\end{eqnarray}
in terms of the associated Legendre polynomials. Here
$\nu \equiv \sqrt{1-k^2}$ is real for super-curvature modes,
$k^2 < 1$, and we have normalized the modes following
Refs.~\cite{HST,Jaume}.
The higher multipoles can be obtained from~\cite{AS}
\begin{eqnarray}\label{Legendre}
P^{1/2}_{\nu-1/2}(\cosh\xi) & = & \sqrt{2\over\pi\sinh\xi}\,
\cosh\nu\xi\,,\\[2mm]
P^{-1/2}_{\nu-1/2}(\cosh\xi) & = & \sqrt{2\over\pi\sinh\xi}\,
{\sinh\nu\xi\over\nu}\,,
\end{eqnarray}
with the recurrence relation
\begin{eqnarray}\label{recurrence}
(\nu^2&-&l^2)P^{-l-1/2}_{\nu-1/2}(\cosh\xi) =
P^{3/2-l}_{\nu-1/2}(\cosh\xi) \nonumber\\[2mm]
&&- (2l-1)\coth\xi\ P^{1/2-l}_{\nu-1/2}(\cosh\xi)\,.
\end{eqnarray}
The scalar harmonics for the first two multipoles can be written as
\begin{eqnarray}\label{Ql01}
Q_{200}(\xi,\Omega) & = & N_0\,\cosh\xi\ Y_{00}(\theta,\phi)\,,\\[2mm]
Q_{21m}(\xi,\Omega) & = & N_1\,\sinh\xi\ Y_{1m}(\theta,\phi)\,,
\end{eqnarray}
where the Klein-Gordon norms $N_0$ and $N_1$ diverge, see
Eq.~(\ref{pikl}). This is not a problem since the only
contribution of these modes to curvature perturbations comes
through the transverse traceless tensor $Q_{ij}$, see
Eq.~(\ref{curva}), and is easy to check that, for $k^2=-3$, it
vanishes identically for the first two multipoles, $l=0$ and $l=1$.
However, it does not vanish for the higher multipoles and thus
the transverse traceless curvature perturbation (\ref{curva})
gets contributions from all the $l\geq2$ multipoles~\cite{Jaume}.
Let us calculate the quadrupole with Eq.~(\ref{recurrence}).
It is clear from this equation that the mode $k^2=-3$, or
$\nu=2$, is very special since the LHS seems to vanish
for $l=2$. However, the RHS also vanishes and
thus we should find the quadrupole in the limit $\nu\to2$,
\begin{equation}\label{P22}
\Pi_{22}(\xi) = \sqrt{24\over\pi}\,{\sinh4\xi - 8\sinh2\xi + 12\xi
\over96\sinh^3\xi}\,.
\end{equation}
The rest of the multipoles can now be obtained from this
expression together with the recurrence relation (\ref{recurrence}).
 
We are interested in the multipole components of temperature
fluctuations induced by this discrete mode.
The best way to analyze its effect on the microwave background
is to study this scalar metric perturbation in the comoving,
uniform-Hubble-constant hypersurface gauge,  in terms of the
gauge invariant potential $\Phi$, as we did in Section III.
 
Let us evaluate the multipole components of the temperature
anisotropies associated with the bubble wall fluctuations. A
formalism that includes supercurvature modes $(k^2<1)$ was
developed in Ref.~\cite{LW}, were it was found that they contribute
to the CMB temperature anisotropies like realizations of a
homogeneous random field. Furthermore, in Ref.~\cite{HST} it
was shown that the $k^2=-3$ mode also corresponds to a
homogeneous random field, once we subtract the non-physical
monopole and dipole contributions.
 
In Section IV, we gave the expression of the angular power
spectrum of the observed temperature fluctuations coming from
the various metric perturbations, see Eq.~(\ref{LCL}). We will
concentrate here on the contribution of the discrete $\nu=2$ mode
associated with the bubble wall quantum fluctuations.
$\langle{\cal R}^2_0\rangle_{\rm wall}$ is the average square
amplitude of the metric perturbation (\ref{Rconst}). The ``window
function'' $I_{\nu l}^2$ indicates how this mode
contributes to the $C_l$'s,
\begin{equation}\label{window}
\nu I_{\nu l} = {1\over5}\,\Pi_{\nu l}(\eta_0) + {6\over5}\,
\int_o^{\eta_0} \Pi_{\nu l}(r)\,F'(\eta_0-r)\,dr\,,
\end{equation}
where $F(\eta)$ is given by Eq.~(\ref{Feta}).
 
It is possible to compute analytically its contribution
to the first multipoles of the angular power spectrum,
for values of $\Omega_0$ close to one,
\begin{eqnarray}\label{first}
C_2 & = & {\pi\over3} \left(8\over25\right)^2 \langle{\cal R}^2_0
\rangle_{\rm wall}\,(1-\Omega_0)^2 \,,\\[2mm]
C_3 & = & {\pi\over15} \left(16\over35\right)^2 \langle{\cal R}^2_0
\rangle_{\rm wall}\,(1-\Omega_0)^3 \,.
\end{eqnarray}
In general we find $C_l \sim \langle{\cal R}_0^2\rangle_{\rm wall}\,
(1-\Omega_0)^l$, and thus the quadrupole dominates over the
rest of the multipoles, like in the case of open universe
Grishchuk-Zel'dovich effect~\cite{OGZ}.
 
For $\Omega_0 < 1$, we have to evaluate numerically the window
functions for the different multipoles, and compute their contribution
to the CMB. In Fig.~1 we show the shape of the angular power
spectrum, normalized to the quadrupole, as a function of multipole
number $l$, for various values of $\Omega_0$. It is clear that none
of them are compatible with a flat spectrum. Note that we recover the
usual Grishchuk-Zel'dovich effect of a dominating quadrupole in the
limit $\Omega_0\simeq1$. In Fig.~2 we show the amplitude 
of the first contributing multipoles of the angular
power spectrum in units of $\langle{\cal R}_0^2\rangle_{\rm wall}$,
as a function of $\Omega_0$. It is clear that for a large range of
$\Omega_0$, the quadrupole dominates the spectrum.
 
In Fig.~3 we show the limits on the curvature perturbation
$\langle{\cal R}_0^2\rangle_{\rm wall}$ as a function of $\Omega_0$,
from the observational limits on $l(l+1)\,C_l < 8\times 10^{-10}$.
We can parametrize this bound as
\begin{eqnarray}
\langle{\cal R}_0^2\rangle_{\rm wall} &\lesssim&
3\times10^{-9}\,,\hspace{1.2cm}
0.1 \lesssim\Omega_0\lesssim 0.4\,,\label{bound1}\\[2mm]
\langle{\cal R}_0^2\rangle_{\rm wall} &\lesssim&
{10^{-9}\over(1-\Omega_0)^2}\,,\hspace{1.1cm}
0.4 \lesssim\Omega_0  \leq 1\,.\label{bound2}
\end{eqnarray}
This is our main result. We will use it to constrain the models of open
inflation in Section IV.
 

\vspace{2cm}
\begin{center}
{\bf FIGURE CAPTIONS}
\end{center}
 
\noindent
{\bf Figure 1:} The shape of the radiation angular power spectra
$l(l+1)C_l$ induced by the $k^2=-3$ mode of bubble wall fluctuations,
normalized to the quadrupole, for the first nine multipoles. The curves
correspond to  $\Omega_0 = 0.2, \, 0.4, \, 0.6, \, 0.8$, from top to
bottom. It is clear that none of them are compatible with a flat
spectrum. Note that we recover the usual Grishchuk-Zel'dovich effect
of a dominating quadrupole in the limit $\Omega_0\simeq1$.
 
\vspace*{24pt}
\noindent
{\bf Figure 2:}  The amplitude of the first three contributing
multipoles, $\,l=2,3,4$ (from top to bottom), of the angular
power spectrum in units of $\langle{\cal R}_0^2\rangle_{\rm wall}$,
as a function of $\Omega_0$. It is clear that for a large range of
$\Omega_0$, the quadrupole dominates the spectrum.

\vspace*{24pt}
\noindent
{\bf Figure 3:}  Limits on $\langle {\cal R}_0^2\rangle_{\rm wall}$,
based on current observational limits on $l(l+1)C_l<8\times10^{-10}$.
The allowed values of $\langle {\cal R}_0^2\rangle_{\rm wall}$ are
those below the curve. In a large range of $\Omega_0$ the quadrupole
provides the strongest constraint.
 
\end{document}